\documentclass[reprint, aps,superscriptaddress,amsmath,amssym,prb]{revtex4-2}

\usepackage{bm}
\usepackage{float}
\usepackage{graphicx}
\usepackage[usenames,dvipsnames]{color}
\usepackage[normalem]{ulem}
\usepackage[svgnames]{xcolor}
\usepackage{bm}
\usepackage{multirow}
\usepackage{titlesec}
\usepackage[utf8]{inputenc}
\usepackage[colorlinks,linkcolor=blue,citecolor=blue,urlcolor=blue]{hyperref}
\usepackage{booktabs}
\usepackage{array}
\usepackage{url}
\begin{document}

\preprint{APS/123-QED}

\title{Nernst effect and its thickness dependence in superconducting NbN films}%

\author{Thomas Bouteiller}
\email{e-mail: thomas.bouteiller@espci.fr}
\affiliation{Laboratoire de Physique et d'\'Etude des Mat\'eriaux \\ (ESPCI - CNRS - Sorbonne Universit\'e)\\ Universit\'e Paris Sciences et Lettres, 75005 Paris, France}

\author{Arthur Marguerite}
\affiliation{Laboratoire de Physique et d'\'Etude des Mat\'eriaux \\ (ESPCI - CNRS - Sorbonne Universit\'e)\\ Universit\'e Paris Sciences et Lettres, 75005 Paris, France}

\author{Ramzy Daou}
\affiliation{Laboratoire de Cristallographie et Sciences des Mat\'eriaux (CRISMAT), Normandie Universit\'e, UMR6508 CNRS, ENSICAEN, UNICAEN, 14000 Caen, France}

\author{Dmitry Yakovlev}
\affiliation{Laboratoire de Physique et d'\'Etude des Mat\'eriaux \\ (ESPCI - CNRS - Sorbonne Universit\'e)\\ Universit\'e Paris Sciences et Lettres, 75005 Paris, France}

\author{St\'ephane Pons }
\affiliation{Laboratoire de Physique et d'\'Etude des Mat\'eriaux \\ (ESPCI - CNRS - Sorbonne Universit\'e)\\ Universit\'e Paris Sciences et Lettres, 75005 Paris, France}

\author{Cheryl Feuillet-Palma }
\affiliation{Laboratoire de Physique et d'\'Etude des Mat\'eriaux \\ (ESPCI - CNRS - Sorbonne Universit\'e)\\ Universit\'e Paris Sciences et Lettres, 75005 Paris, France}

\author{Dimitri Roditchev}
\affiliation{Laboratoire de Physique et d'\'Etude des Mat\'eriaux \\ (ESPCI - CNRS - Sorbonne Universit\'e)\\ Universit\'e Paris Sciences et Lettres, 75005 Paris, France}

\author{Beno\^it Fauqu\'e}
\affiliation{JEIP  (USR 3573 CNRS), Coll\`ege de France \\ Universit\'e Paris Sciences et Lettres, 75005 Paris, France}

\author{Kamran Behnia}
\email{e-mail: kamran.behnia@espci.fr}
\affiliation{Laboratoire de Physique et d'\'Etude des Mat\'eriaux \\ (ESPCI - CNRS - Sorbonne Universit\'e)\\ Universit\'e Paris Sciences et Lettres, 75005 Paris, France}

\date{\today}

\begin{abstract}
Superconducting thin films and layered crystals display a Nernst signal  generated by short-lived Cooper pairs above their critical temperature. Several experimental studies have broadly verified the standard theory  invoking Gaussian fluctuations of a two-dimensional superconducting order parameter. Here, we present a study of the Nernst effect in granular NbN thin films with a thickness varying from 4 to 30 nm, exceeding the short superconducting coherence length and putting the system in the three-dimensional limit. We find that the Nernst conductivity decreases linearly with reduced temperature ($\alpha_{xy}\propto \frac{T-T_c}{T_c}$), but the amplitude of $\alpha_{xy}$ scales with thickness. While the temperature dependence corresponds to what is expected in a 2D picture, scaling with thickness corresponds to a 3D picture. We argue that this behavior indicates a 2+1D situation, in which the relevant coherence length along the thickness of the film has no temperature dependence. We find no visible discontinuity in the temperature dependence of the Nernst conductivity  across T$_c$. Explaining how the response of the superconducting vortices evolves to the one above the critical temperature  of short-lived Cooper pairs emerges as a challenge to the theory.
\end{abstract}

\maketitle

\section{Introduction}
The Nernst effect \cite{Behnia2016} refers to the emergence of a transverse component in the thermoelectric response in the presence of  a magnetic field. Suppose that a thermal gradient is applied  along the x-axis and the magnetic field is oriented along the z-axis, if this leads to an electric field along the y-axis, there is a Nernst signal defined as:
\begin{equation}
N=\dfrac{E_y}{\nabla_xT}
\end{equation}

This quantity, expressed in $\rm \mu V/K$, is the one directly accessible by the experiment. But the following pair of equations is often more interesting:

\begin{align}
\label{eqn:eqlabel}
\begin{split}
\vec{J}_e=\sigma \vec{E}- \alpha \vec{\nabla}T ,
\\
\vec{J}_q=\alpha T \vec{E}- \kappa \vec{\nabla}T . 
\end{split}
\end{align}
\begin{figure}[ht!]
\includegraphics[width=0.95\linewidth]{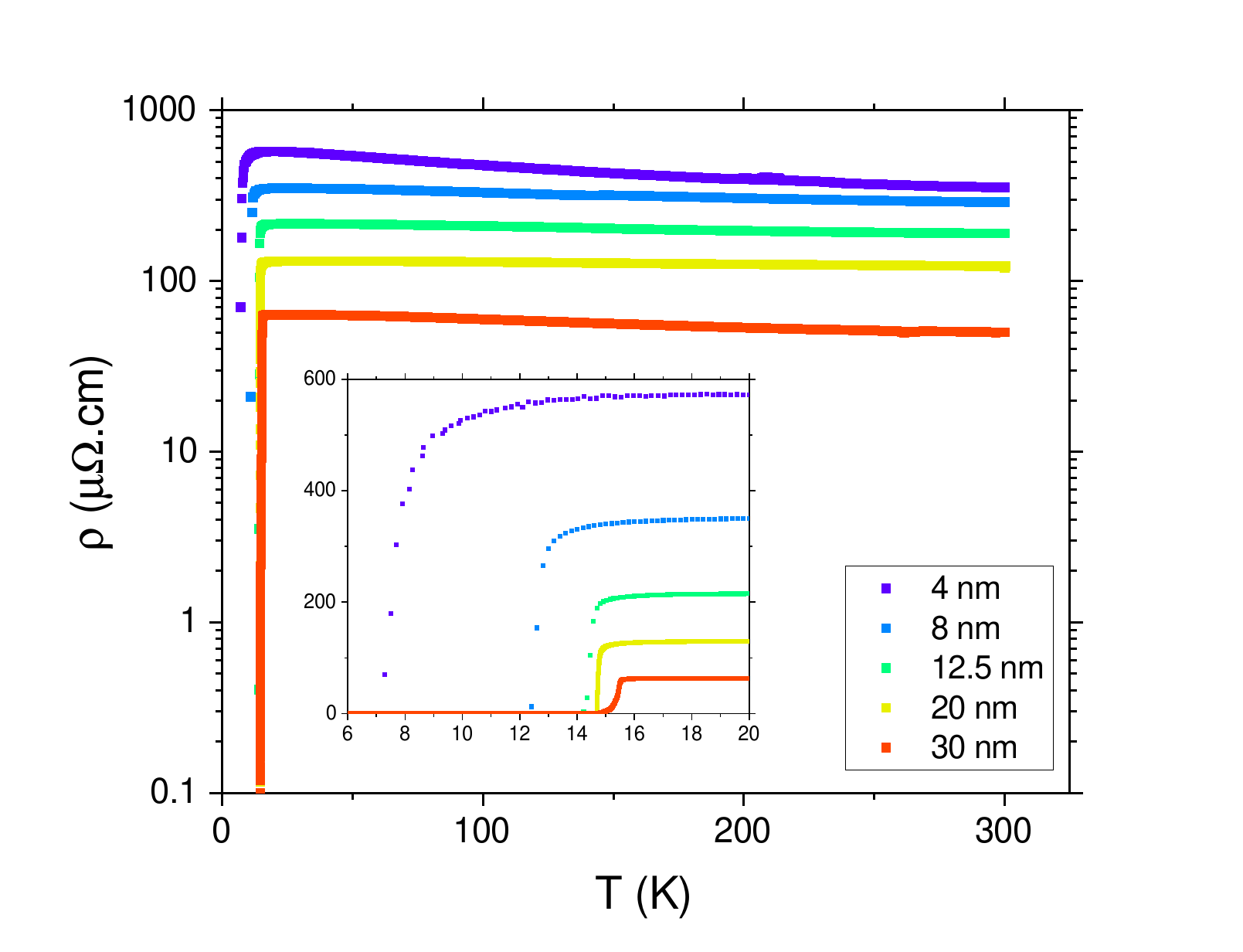}
\centering
\caption{Temperature dependence of the electrical resistivity of the five samples used in this study, in a semi-log plot. With decreasing thickness, the amplitude of normal state resistivity increases and the temperature dependence, a mild decrease with cooling due to weak localization becomes more pronounced. The inset  shows the data in a linear plot near the superconducting transition. One can see that resistivity in thinner samples starts to decrease above the transition, as consequence of enhanced fluctuations. }
\label{Fig:1}
\end{figure}

$\vec{J}_e$ and $\vec{J}_q$ are flux densities of electric charge and thermal energy, which relate to the electric field ($\vec{E}$) and  temperature gradient ($\vec{\nabla}T$) vectors through electric, $\sigma$,  thermal $\kappa$, and thermoelectric, $\alpha$,  conductivity tensors. A finite Nernst signal implies a finite off-diagonal component of $\alpha$. The two quantities are linked by:

\begin{equation}
\alpha_{xy}=N \sigma_{xx}-S\sigma_{xy}
\label{def-alpha}
\end{equation}

In this relation, the second term can be often neglected in the zero-field limit. Then, $\alpha_{xy}$ becomes simply the Nernst signal divided by the resistivity. Annoyingly,  $\alpha_{xy}$, which in three dimensions is expressed in $A/(K.m)$ and in A/K in two dimension, has not acquired  a standardized name in scientific literature. Among others, it has been called the transverse Peltier coefficient  or the transverse thermoelectric conductivity (despite the fact that the units of Peltier coefficient are volts). Throughout this paper, we will simply call it the Nernst conductivity.

Two decades ago, following the experimental observation of a Nernst signal above the critical temperature in cuprates \cite{Wang2001,Wang2006}, Ussishkin \textit{et al.} \cite{Ussishkin} derived the following theoretical expression for the amplitude of the Nernst conductivity due to the Gaussian fluctuations in the normal state of a two-dimensional  superconductor: 

\begin{equation}
    \alpha_{xy}^{2D}=\dfrac{1}{6\pi}\dfrac{k_Be}{\hbar}\dfrac{\xi^2}{l_B^2}
    \label{USH-2D}
\end{equation}

This simple expression links the amplitude of Nernst conductivity, to the quantum of thermoelectric conductance ($\frac{k_Be}{\hbar}$) \cite{Behnia2015b}, the magnetic length and a single material-dependent parameter (the superconducting correlation length $\xi$). 

Soon, an experiment confirmed what is expected according to Equation in an amorphous 2D superconductor \cite{Pourret2006}. This first confirmation was followed by other experimental studies,\cite{Pourret2007,Spathis2008,Chang2012,Tafti2014,Cyr2018,jotzu2021,XinQiLi2020} verifying the relevance of  Eq.  \ref{USH-2D} to other superconductors. On the theoretical side, several studies \cite{Michaeli2009,Serbyn2009,Levchenko2011,Glatz2020} confirmed  and expanded the initial prediction by Ussishkin \textit{et al.} \cite{Ussishkin}. It is fair to say that the thermoelectric response due to fluctuating superconductivity is well understood in its broad lines (See \cite{Pourret2009,Behnia2016} for reviews).

\begin{figure*}[htb!]
\includegraphics[width=16cm]{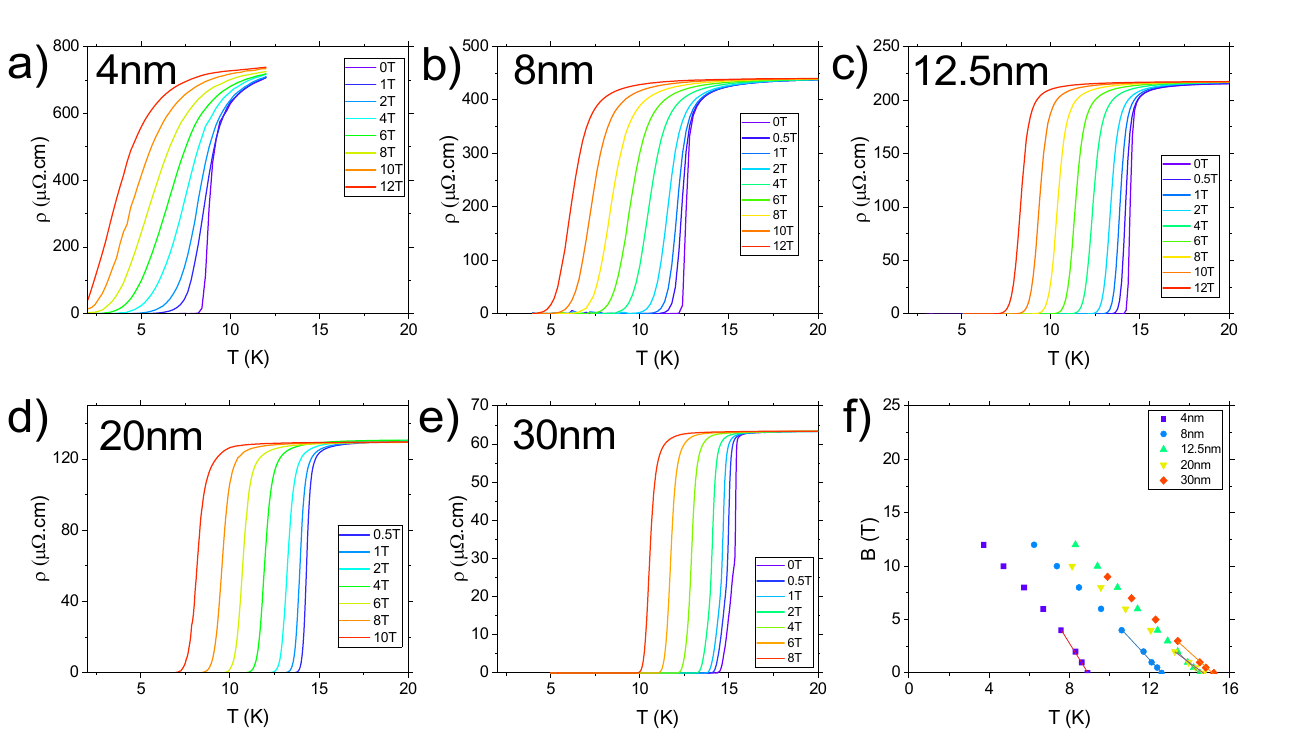}
\caption{\label{fig:TcvsB} Temperature dependence of the electrical resistivity of the five samples  in the presence of the magnetic field (a-e). Critical temperature decreases with increasing magnetic field. However, superconductivity is not fully suppressed by 12 T, our largest magnetic field. f) The variation of T$_c$ with magnetic field. The slope at low fields ($\le$4T) was used to determine the  upper critical field at zero temperature, $H_{c2} (0)$, and the  superconducting coherence length $\xi_0$.  }
\end{figure*}

On the other hand, the existence of a Nernst signal below the superconducting critical transition has been known for a long time \cite{LOWELL1967,solomon1967,HUEBENER20211353975,Huebener_1995,Li1994}. Its origin has been tracked down to the motion of superconducting vortices upon the application of a temperature gradient \cite{stephen1966,Maki1968,Maki1971,HUEBENER1967947,Sergeev_2010}. Ironically, our understanding of this signal has gone along an opposite trajectory and regressed in recent years \cite{Rischau2021}. A rigorous account of the amplitude of the experimentally observed Nernst signal in the vortex liquid state, and in particular of its link to the vortex transport entropy, is missing \cite{Behnia_2023,Hu2024,Ienaga2024}. A consensus is yet to be found  \cite{guo2024amplitudevortexentropysemiclassical}.

In this paper we present an extensive study of the  Nernst effect across the superconducting critical temperature in NbN thin films. In its crystalline form, NbN has a sodium chloride structure and a critical temperature as high as 17.8 K. It has been studied by various probes, as a crystal \cite{Chen2005b} and, much more frequently, as a thin film \cite{Chokal2008,SEMENOV,Kamlapure2010,Mondal2011, Yong2013,Noat2013}.  Strongly disordered NbN thin films (with a thickness of 3 nm) display a Berezinskii-Kosterlitz-Thouless transition, where the zero-field resistive transition is smeared by phase fluctuations and the proliferation of vortices and anti-vortices \cite{Mondal2011b,Weitzel2023}.  

NbN thin films are a compelling platform for a study of Nernst effect, given that its  critical temperature is much higher than other thin film superconductors subject to a Nernst study near a superconductor-insulator transition (such as Nb$_{1-x}$Si$_x$ \cite{Pourret2006,Pourret2007}, InO$_x$ \cite{Spathis2008,Roy2018} and MoGe \cite{Rischau2021,Ienaga2024}). The rather high critical temperature allowed us to track the evolution of the Nernst effect both in the vortex state and in the fluctuating regime. We studied the Nernst signal  up to  twice the critical temperature, between 0.5 and 12 T and in samples with a thickness  ranging from 4 to 30 nm.  We found that, above the critical temperature, the amplitude of  $\alpha_{xy}$ scales with the thickness of the sample, indicating a 3D behavior. Nevertheless, the temperature dependence of $\alpha_{xy}$ follows what is expected in the  the 2D picture. We argue that this is because the size of the fluctuations along the magnetic field axis is not set by the coherence length but by  the sample thickness. We also found that $\alpha_{xy} (T)$ does not show a clear discontinuity across the critical temperature. This indicates that there is a smooth transition between the regime in which the signal is generated by vortices and the one where it is produced by superconducting fluctuations.

\section{Samples and methods}
NbN thin films were deposited by DC-magnetron sputtering on amorphous (fused) SiO$_2$ substrates, using a Nb target in a nitrogen/argon atmosphere. The substrate was heated at 800 C during deposition. We chose this substrate because of its low thermal conductivity, which allowed us to easily obtain sizable temperature gradients in our temperature range of interest. The thickness of the samples was estimated from the calibration of deposition source and confirmed by comparing with previously reported properties studies \cite{Kern_2024,Jing_2023, SEMENOV}. 

Electric and thermoelectric conductivity measurements were performed using a Physical Properties Measurement System (PPMS). We first measured the thermal conductance of our amorphous substrate from 2 K to 50 K, and found a good agreement with was previously reported \cite{Zeller1971}.  Then, for each sample, the Nernst signal was measured using a small resistive chip as a heater, which generated  a thermal gradient. A bare-chip Cernox temperature sensor was glued to the sample and used to measure the temperature of the sample. The temperature gradient was obtained from the amplitudes of the heat current and the previously calibrated substrate thermal conductance. In order to stay in the linear response regime, we used low heating currents keeping $\Delta T/T$ smaller than 10$\%$. Electrical contacts were made using 4929N silver paste. Transverse voltage was obtained by measuring Nernst voltage for positive and negative fields, and antisymmetrizing the obtained voltage $V_{xy}=0.5*(V(B)-V(-B))$ to eliminate parasitic longitudinal signals.

The properties of the five samples studied in this work are summarized in table \ref{table:1}, and the temperature dependence of their electrical resistivity is shown in Figure \ref{Fig:1}. In all samples, resistivity increases with decreasing temperature, as expected in a dirty metal. As the thickness is reduced, resistivity increases  from 70 $\rm \mu\Omega.cm$ at 30 nm to 580 $\rm \mu\Omega.cm$ at 4 nm. Note that while the resistivity increases by a factor 8, the sheet resistance increases by a factor $>$50. Concomitantly, the critical temperature decreases from 15.5 K to 8.9 K.  Both these features were reported in previous studies on NbN thin films. They are generic to disordered superconducting thin films \cite{FINKELSTEIN1994636,Gantmakher_2010,Ummarino_2025}.  Even in our thinnest sample is not an insulator. Its sheet resistance is well below $h/4e^2=6.45 \rm k\Omega$, and the normal state resistivity displays a mild temperature dependence, characteristic of weak localization in a dirty metal.

\section{Results}
\begin{table*}
\centering
\begin{tabular}{c c c c c c c}
\hline Sample& Thickness (nm) & $R_{S,max}$ ($\rm \Omega/sq$)& $\rho_{max} \rm(\mu\Omega.cm)$& RRR &$T_c$ (K)&   $\xi_0^{Hc2}$ (nm) \\ 
\hline
\hline1& 4& 1450& 580&0.52 &  8.7& 4.3 \\  
2&8&435 &350&0.82 & 12.4  &  4.4  \\  
3&12.5 &175&220&0.86& 14.2 & 4.4 \\ 
4&  20&65& 130&0.92 &14.7 &  4.8 \\ 
5& 30&25 &70&0.85 &15.5  &4.3\\ 
\hline
\end{tabular}
\caption{Properties of the five NbN thin films  used in this study. $\rho_{max}$ is defined as the maximum of resistivity above the superconducting transition. The residual resistivity ratio RRR is defined as $\rho$(300K)/$\rho_{max}$}

\label{table:1}
\end{table*}
\begin{figure}[ht!]
\centering
\includegraphics[width=1\linewidth]{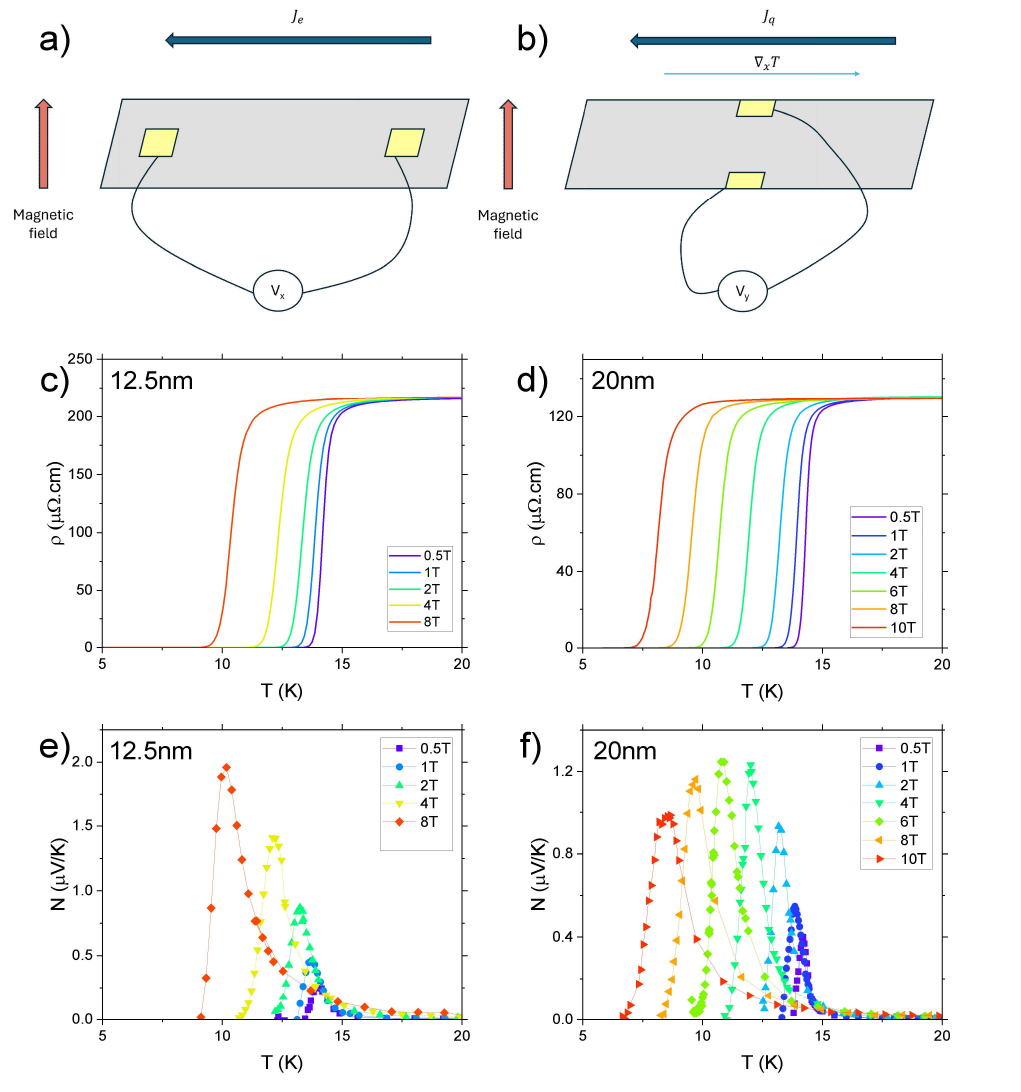}
\caption{\label{fig:NvT} Experimental configurations for measuring the electric resistivity (a) and the Nernst signal  (b). The temperature dependence of $N$ and $\rho$ in the 12.5 nm sample (c,e)  and in the 20 nm sample (d, f). In both samples, at low temperature, when there is no resistivity,  the Nernst signal is zero. It peaks during the transition and attains an amplitude of the order of the $\rm \mu V/K$ and then  decreases but remains finite above the critical temperature. Note the non-monotonic variation of the amplitude of the Nernst peak  with magnetic field in the 20 nm sample.}
\end{figure}

\subsection{The upper critical field}
Fig \ref{fig:TcvsB} shows the evolution of the resistive superconducting transition with applied magnetic field.  In all samples, as the magnetic field increases, the critical temperature decreases  and the transition becomes wider. Our maximum field of 12 T was not sufficiently strong to suppress superconductivity at 2 K in any of the samples.  

\begin{figure*}[ht!]
\centering
\includegraphics[width=17cm]{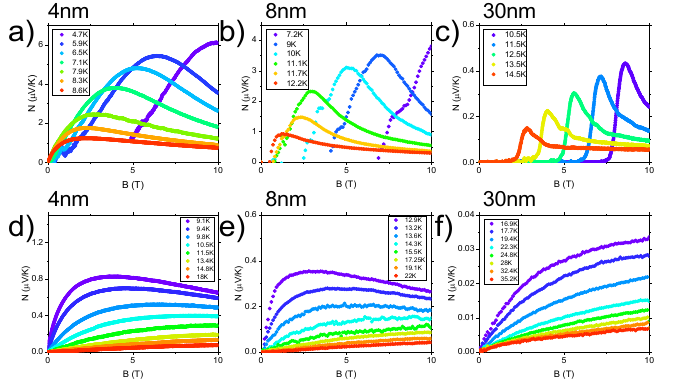}\caption{\label{fig:NvB} The Nernst signal as a function of the magnetic field below  (a-c) and above (d-f) the critical temperature in three samples with different thicknesses. The amplitude of $N$ decreases with increasing thickness. Note the warming-induced shift of the  peaks in opposite directions below and above T$_c$.}
\end{figure*}

Fig \ref{fig:TcvsB}f  displays how the critical temperature changes as a function of the magnetic field. We used Werthamer–Helfand–Hohenberg theory \cite{Werthamer1966} and our $H_{c2} (T)$ data to estimate the zero-temperature upper critical field,  $H_{c2} (0)$ : 
\begin{equation}
    H_{c2}(0)=-0.69T_c\dfrac{dH_{c2}}{dT}   
\end{equation}

This allowed us then to quantify the coherence length at zero temperature using :
\begin{equation}
    \xi_0=\sqrt{\dfrac{\phi_0}{2\pi H_{c2}(0)}}
\end{equation}

The results  are shown table \ref{table:1}. Values of coherence obtained with this method will be written $\xi_0^{Hc2}$. Remarkably, despite an almost twofold variation of the critical temperature, the zero-temperature coherence length is almost the same. Note that for all studied samples, $\xi_0 \sim$ 4.5 $\pm$ 0.3 nm, that is shorter than the thickness of all samples, save for the thinnest one.

\subsection{Concomitance between the Nernst response and the electric resistivity}
Fig \ref{fig:NvT} shows the Nernst and the resistivity data in two samples. The experimental configurations are sketched in panels a and b. Panels c and e show the temperature dependence of $N$  and $\rho$ across the superconducting transition in the  12.5 nm thick sample. Panels d and f show similar data for the 20 nm thick sample. In both samples, there is a visible correlation between the two sets of data. At each magnetic field, at the lowest temperature, when there is no electric resistivity, $N (T)$ is zero. It becomes finite when  dissipation begins. It peaks and decreases afterwards, concomitant with the resistive superconducting transition. Such a behavior has been seen in numerous superconductors \cite{HUEBENER1967947,Li1994, Logvenov1997,Gollnik1998,Pourret2011,XinQiLi2020,Rischau2021}. Below the critical temperature, vortex displacement is the source of dissipation and is assumed as the generator of the finite Nernst response in the superconducting state. 

 Above  the critical temperature, the amplitude of $N$ decreases drastically, but remains finite. Note also that the peak Nernst effect decreases with increasing thickness, and therefore with the amplitude of the resistivity. It decreases from 2 $\rm \mu V/K$ in the  12.5 nm sample to  1.3 $\rm \mu V/K$ in the 20 nm one. This is a $\approx$ 1.54 times decrease, comparable to the $\approx$ 1.6 decrease in resistivity.  We will discuss this in more detail below. Note, however,  that the order of magnitude of this vortex Nernst peak is a few $\mu V/K$ in all samples, and comparable to what was  found in many superconductors with critical temperature varying from a few hundreds of mK to 100 K. This is in sharp contrast with  metals, where the quasi-particle Nernst response can have an amplitude ranging from  nV/K to a few mV/K \cite{Behnia2016}. 

\subsection{The field dependence and the ghost critical field}
Fig \ref{fig:NvB} shows the field dependence  of the Nernst effect in three (4, 8 and 30 nm) samples at different temperatures.  The upper (lower) panels show the data below (above) the  transition temperature. In both cases the signal is non-monotonic below and above $T_c$. It increases first, reaches a maximum  before decreasing. Note the difference in the units of the vertical axes in the lower and upper panels.  Above the critical temperature, the amplitude of the signal is lower by one order of magnitude. Note the contrast in the way the peaks shift as a function of temperature.  In the superconducting state (upper panels), with decreasing temperature, the peak shifts to a higher field.  In the normal state (lower panels), the displacement is in the opposite direction. 

This contrast, first observed in thin films of Nb$_{0.15}$Si$_{0.85}$ \cite{Pourret2006}, was also observed in a hole-doped \cite{Chang2012} and an electron-doped  cuprates \cite{Tafti2014}. Its origin was tracked down\cite{Pourret2007,Pourret2009} to the presence of a "ghost critical field" \cite{Kapitulnik_1985} in the normal state, mirroring the temperature dependence of the upper critical field of the superconductor. Both these field scales vanish at the critical temperature. Each of them is locked to the correlation length, $\xi(T)$, at one side of the transition.  The field dependence of the Nernst response above and below T$_c$ reflects the contrasting temperature dependence of $\xi(T)$  in the two sides.

The  peak amplitude of the Nernst signal in the vortex state displays a significant thickness dependence.  It decreases from 6.2 $\rm \mu V/K$ at 4 nm to only  $0.4 \rm \mu V/K$ at 30 nm. We will see below that this is a consequence of the relevance of the third dimension. 

\begin{figure}[h!]
\includegraphics[width=0.9\linewidth]{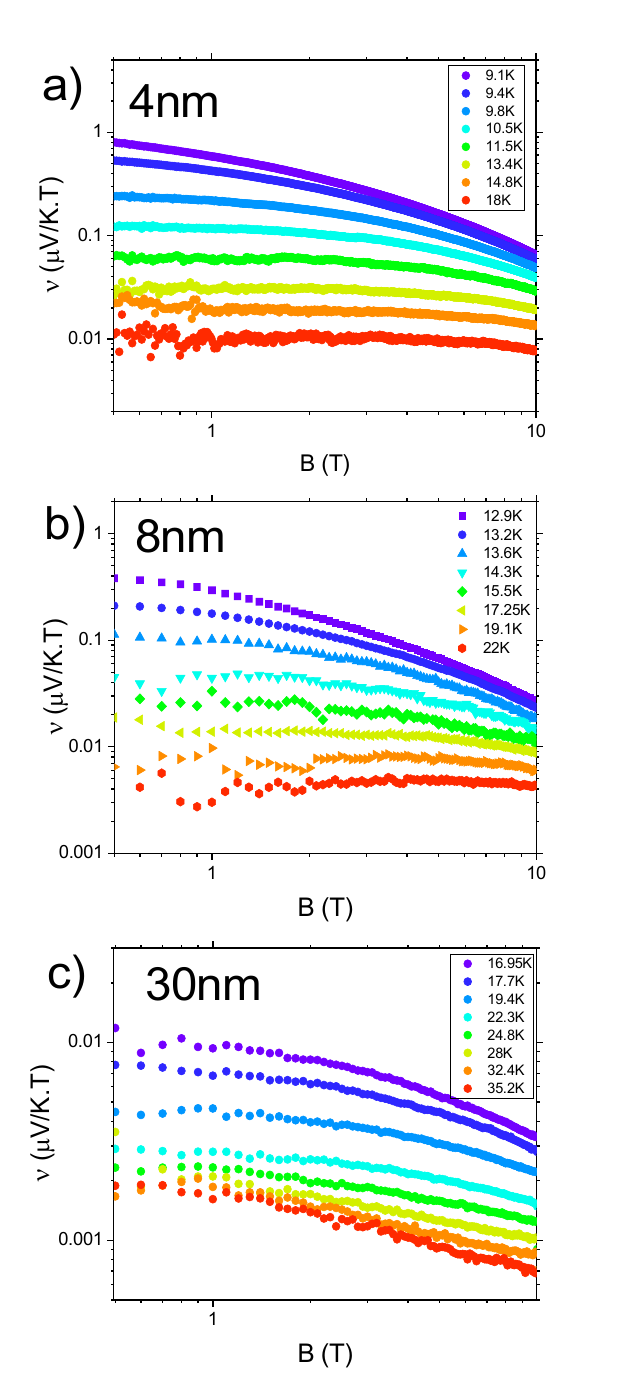}
\caption{\label{fig:nuvb} The Nernst coefficient, $\nu=N/B$ above the critical temperature plotted versus magnetic field for a) 4 nm b) 8 nm and c) 30 nm samples. At the low  field limit, $\nu$ is constant and is solely set by the temperature. At 10 T, curves start to collapse on one another.  A regime where $\nu$ is temperature independent and only set by the magnetic field starts to emerge. }
\end{figure}

\subsection{The Nernst coefficient in the normal state}

Fig \ref{fig:nuvb} shows the field dependence of the Nernst coefficient, $\nu=N/B$, above the critical temperature in three (4, 8 and 30 nm) samples in a log-log plot.

As first reported by Pourret \textit{et al.} \cite{Pourret2006,Pourret2007}, such plots can reveal that the field and the temperature dependence of $\nu$ depend on the relative amplitude of the magnetic length ($\ell_B=\sqrt{\frac{\hbar}{eB}}$)  and the correlation length $\xi=\xi_0\sqrt{\frac{T_c}{T-T_c}}$.  They can highlight two distinct regimes. At low fields, when the magnetic length is longer than the superconducting coherence length, $\nu$  depends on temperature but not on magnetic field. At high magnetic fields, on the other hand, the magnetic length becomes much shorter than the coherence length, and all curves start to collapse on top of each other. This is a regime where the amplitude of $\nu$ is set by the magnetic field (and not by temperature). Experimentally, such a behavior was observed in both Nb$_{0.15}$Si$_{0.85}$ \cite{Pourret2006} and in Eu-doped La$_{1-x}$Sr$_x$CuO$_4$ \cite{Chang2012}. 

Comparing our Fig \ref{fig:nuvb}, with Figure 8 in ref. \cite{Behnia2016},  one can detect in our data a low-field regime, in which $\nu$ becomes independent of magnetic field. On the other hand, the high-field regime  (where $\nu$ becomes temperature-independent) is not attained in our measurements. It requires a magnetic field significantly larger than 10 T. Nevertheless, a tendency towards it is detectable.

\begin{figure*}[htp!]
\includegraphics[width=1\linewidth]{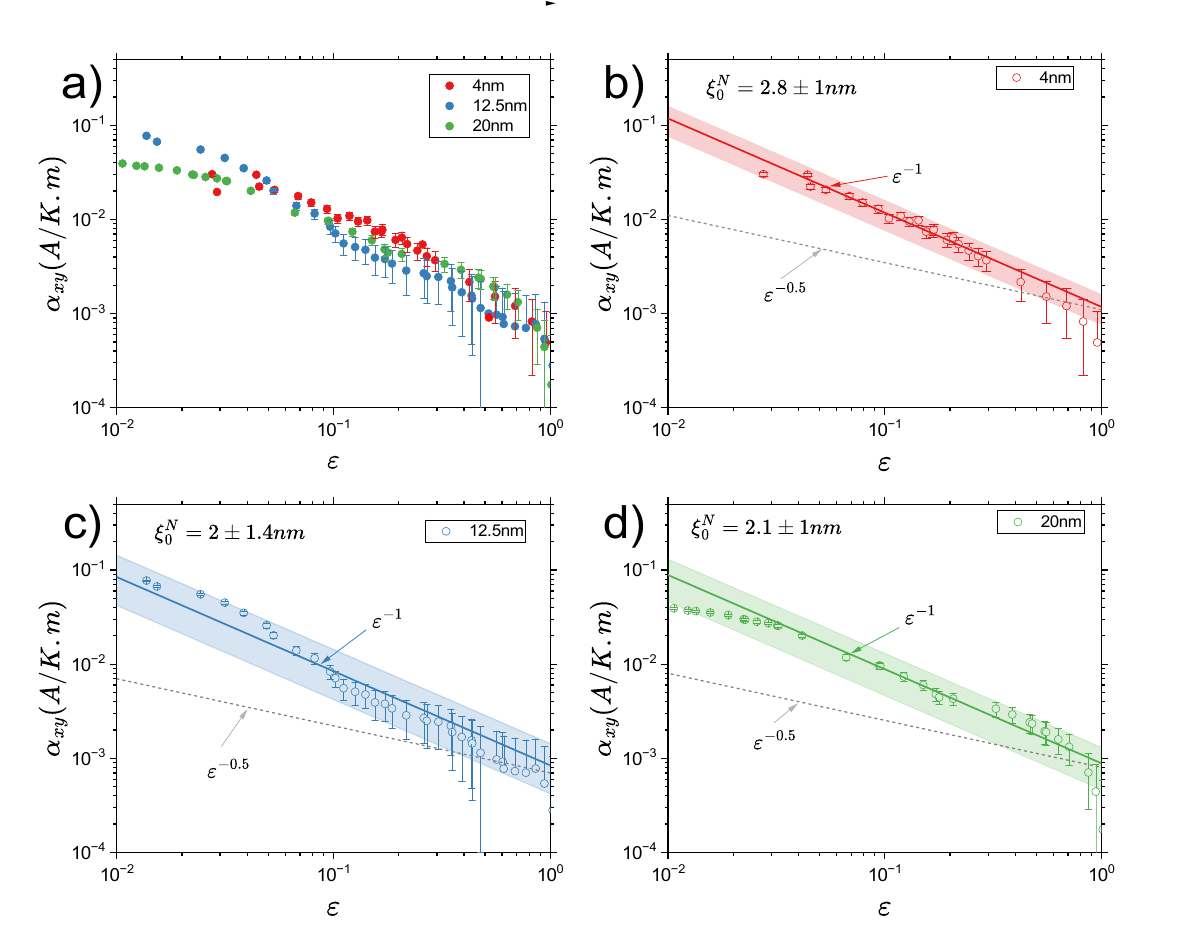}
\caption{\label{fig:alphalowB} a) The Nernst conductivity, $\alpha_{xy}$, extracted from the Nernst and the resistivity data at 0.5 T, plotted  against the reduced temperature $\varepsilon=\frac{T-T_c}{T_c}$ for three samples up to twice the critical temperature. The three curves (expressed in A/K.m  ) are on top of each other, displaying a three-dimensional behavior. The same data is plotted  for each sample in panels b), c) and d). The solid line indicates a temperature dependence of $\varepsilon^{-1}$, corresponding to Equation \ref{USH2+1}. The gray dashed line indicates a temperature dependence of $\varepsilon^{-0.5}$ as expected by Equation \ref{USH-3D}.}
\end{figure*}

\subsection{The amplitude and the temperature dependence of the Nernst conductivity}
The combination of resistivity and Nernst data allows us to determine the amplitude of $\alpha_{xy}$ in the samples.  Fig \ref{fig:alphalowB} shows $\alpha_{xy}$ (that is, $\frac{N}{\rho}$) in three (4, 12.5 and 20 nm) samples at 0.5 T, as a function of the reduced temperature, $\varepsilon$=(T-T$_c$)/T$_c$.    Here, $T_c$ is taken as the midpoint of the resistive superconducting transition. A field of 0.5 T is enough to reach the low field regime discussed previously, and $\nu$ only depends on the temperature. Remarkably, in a significant range of reduced temperature ($0.05<\varepsilon<1$), the three curves fall on one another. Using the 2D definition of $\alpha^{2D}_{xy}=N/R_S$, expressed in A/K, would have yielded a sample dependent Nernst conductivity, varying from 0.04 A/K at 4 nm to 0.2 A/K at 20 nm, a five fold increase.  As seen in the figure, in this temperature range, which extends to twice the critical temperature, $\alpha_{xy}\propto \varepsilon ^{-1}$. Given that $\xi^2\propto\varepsilon ^{-1} $, this is, indeed,  what is expected by Equation \ref{USH-2D}.

\section{Discussion}
Our results make NbN yet another superconducting platform in which the Gaussian superconducting fluctuations theory\cite{Ussishkin,Michaeli2009,Serbyn2009,Levchenko2011,Glatz2020} has been tested. The list of such superconductors consists of thin films of Nb$_{0.15}$Si$_{0.85}$ \cite{Pourret2006}, cuprate crystals (both electron-doped \cite{Tafti2014} and hole-doped \cite{Kokanovic2009,Chang2012}), thin films of Nb-doped SrTiO$_3$ \cite{Rischau2021} and  a fullerene superconductor (compressed powder K$_3$C$_{60}$ \cite{jotzu2021}). In all these cases, a Nernst signal was detected far above the critical temperature and its temperature dependence was found to follow what is expected by Equation \ref{USH-2D}.  

\subsection{Dimensionality of the Gaussian fluctuations}
In their seminal paper on the Nernst response caused by Gaussian fluctuations, Ussishkin,  Sondhi and Huse,  in addition to Equation \ref{USH-2D}, derived the following expression for the three dimensionnal case \cite{Ussishkin}:
\begin{equation}
    \alpha_{xy}^{3D}=\dfrac{1}{12\pi}\dfrac{k_Be}{\hbar}\dfrac{\xi}{l_B^2}
    \label{USH-3D}
\end{equation}

\begin{table*}
\centering
\begin{tabular}{c c c c c c }
\hline Compound & $T_c$ (K) &$H_{c2}(0)$ (T)&   $\xi_0^{Hc2}$ (nm)& $\xi_0^N$ (nm) &reference \\ 
\hline
\hline  NbN  &8.7-14.7 &17  &4.4 & 2.3 & This work\\
$\rm La_{1.69}Eu_{0.2}Sr_{0.11}CuO_4$& 3.86 &6 &  7&3.8& \cite{Tafti2014}\\ 
$\rm Pr_{1.83}CeO_{0.17}CuO_4$ &19.5& 3 & 10&14&\cite{Chang2012} \\ 
$\rm Nb_{0.15}Si_{0.85}$ &0.38 & 1.1  &  17&10&\cite{Pourret2006} \\  
$\rm SrTi_{0.99}Nb_{0.01}O_3$& 0.32 &  0.085& 62&52&\cite{Rischau2021} \\  
\hline
\end{tabular}
\caption{Critical temperature, upper critical field, coherence length $\xi_0^{Hc2}$ obtained from H$_{c2}$, and coherence length $\xi_0^N$ deduced from the Nernst effect amplitude in superconductors in which  Gaussian fluctuations theory has been verified.}

\label{table:3}
\end{table*}

In both equations, the Nernst conductivity is proportional to the quantum of thermoelectric conductance, $k_be/\hbar$, and the only sample dependent parameter is the coherence length. In both equations, the temperature dependence is set by the ratio of $\xi$ and $\ell_B$. However, the difference in exponents (reflecting the difference in units (A/K in 2D, and A/K.m in 3D), leads to a qualitatively different temperature dependence: $\alpha_{xy}^{2D} \propto \varepsilon^{-1}$ and $\alpha_{xy}^{3D} \propto \varepsilon^{-1/2}$.

As we saw above, $\alpha_{xy}$ is three-dimensional in our NbN films.  Fig.\ref{fig:alphalowB} shows that it scales with the thickness of the film. Moreover, as discussed previously, a 2D model would yield a Nernst conductivity 5 times higher in the 20 nm sample than in the 4 nm, indicative of a coherence length more than twice higher in the 20 nm, as $\alpha^{2D}_{xy} \propto \xi_0^2$. This is in contradiction with the coherence length obtained from transport measurement. On the other hand, its temperature dependence does not follow $\varepsilon^{-1/2}$ but $\propto \varepsilon^{-1}$, which is what is expected in two dimensions.

Faced with this apparent paradox, we note first that our situation is neither two-dimensional (the thickness exceeds the coherence length), nor truly three-dimensional (the thickness is not infinite and there is therefore confinement along the c-axis).  Now, as seen in equation \ref{def-alpha}, $\alpha_{xy}$ is the product of the Nernst signal, (which does not depend on dimensionality), and conductivity (which does). The theoretical 3D  expression (Equation \ref{USH-3D}) is derived assuming a stack of 2D layers each $2\xi$ thick. It is the temperature dependence of this thickness which changes the exponent of the power law from -1 (in Equation \ref{USH-2D}) to -1/2  (in Equation \ref{USH-3D}). We need to reconcile the relevance of three-dimensionality of $\alpha_{xy}$, which requires the introduction of a relevant length scale along the c-axis, with the existence of a bound on this length scale. Now, since $\xi(T)$  diverges at the critical temperature, the easiest way out is to replace $2\xi(T)$ with $2\xi_0$, which has no temperature dependence. In this case, Equation \ref{USH-2D} becomes :

\begin{equation}
    \alpha_{xy}^{(2+1)D}=\frac{1}{12\pi}\frac{k_Be}{\hbar}\frac{\xi^2}{l_B^2} \frac{1}{\xi_0}
    \label{USH2+1}
\end{equation}

This equation yields an $\alpha_{xy}$ with A/K.m units and a linear dependence on the inverse of reduced temperature. The zero-temperature coherence length,  $\xi_0^N$, derived by this expression is $2.8\pm1$ nm for the 4nm sample, $2\pm1$ nm for the 12.5 nm sample, and $2.1\pm1$ nm for the 20 nm sample, to be compared with $\xi_0^{Hc2}\approx 4$ nm, derived from the upper critical field extrapolated to zero temperature.  The two other options would be to introduce thickness or the lattice parameter as the relevant length scales. However, none of them gives a reasonable account of the experimental amplitude of $\alpha_{xy}$ and its insensitivity to thickness. 

Table \ref{table:3} compares our results with was previously reported in superconductors in which $\alpha_{xy}$ has been quantified \cite{Rischau2021,Chang2012,Tafti2014,Pourret2006}. It lists the critical temperature, the upper critical field and the superconducting coherence length obtained by two distinct methods. The coherence length in NbN, fifteen times shorter than in $\rm SrTi_{0.99}Nb_{0.01}O_3$ (62nm), is the shortest in the whole list. This makes its $\alpha_{xy}$ the smallest.  Note that in our case, determination of the amplitude of $\xi_N$ depends on the numerical prefactor in Equation \ref{USH2+1}, which is yet to be rigorously determined. 

\subsection{Continuity of $\alpha_{xy}$ at the critical temperature}
\begin{figure}[htbp]
\includegraphics[width=1\linewidth]{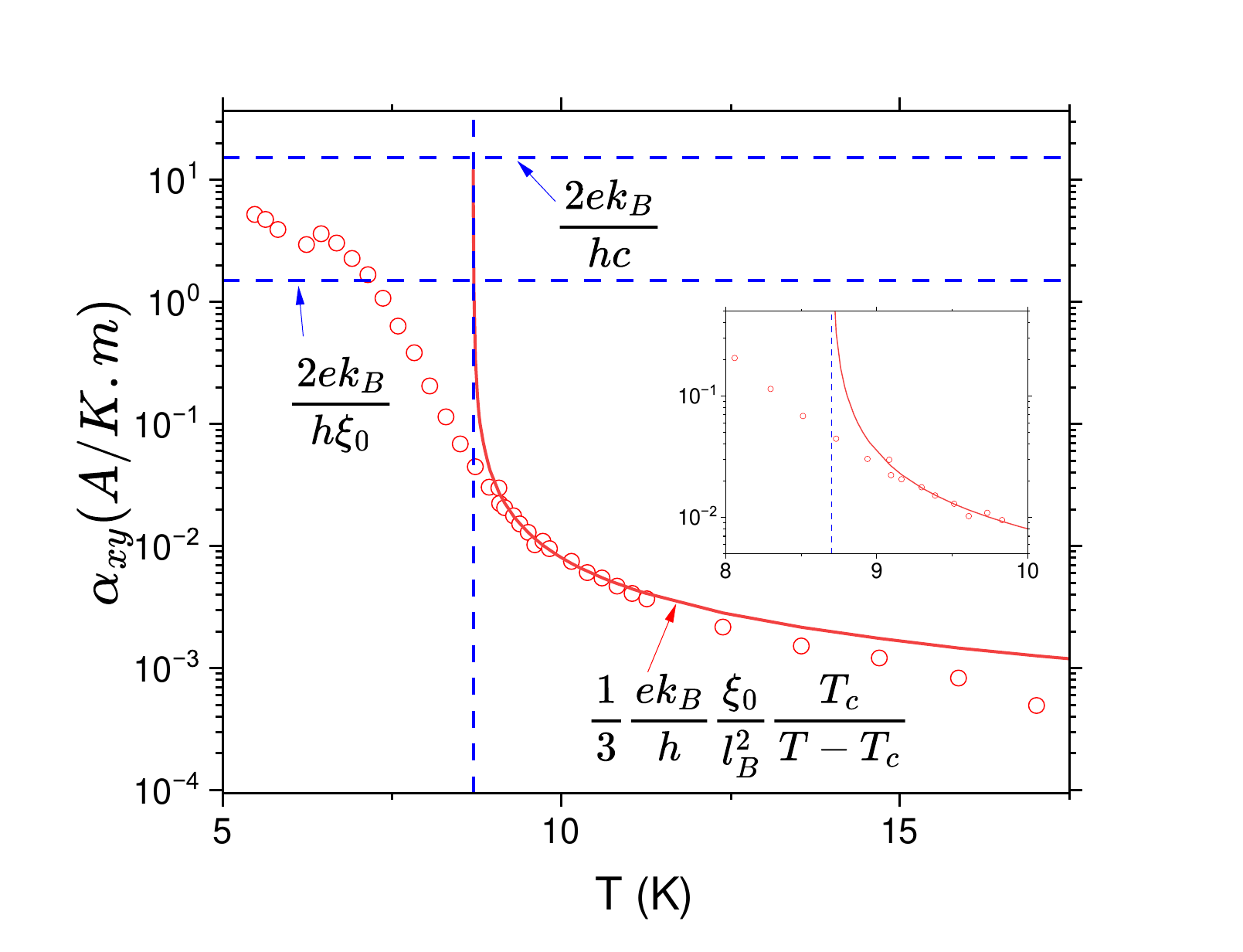}
\caption{\label{fig:alpha} Temperature dependence of $\alpha_{xy}$ (at 0.5T) in the 4 nm sample across the superconducting transition. The solid red line indicates the USH equation using our modified 2+1D model with $\xi_0$=2.8nm. The vertical dashed line marks the critical temperature at zero field. The horizontal dashed line represents the ratio of Boltzmann constant $k_B$ and a magnetic flux of h/2e, divided by the lattice constant c=0.44nm, 15.1 A/K.m, and divided by $\xi_0$, 1.5 A/K.m}
\end{figure}

As seen in Fig.\ref{fig:alphalowB}, close to the critical temperature, that is when $\epsilon$ becomes very small, $\alpha_{xy}$ in all samples deviates from what is theoretical expected. According to  Equation \ref{USH2+1}, it should diverge at the critical temperature. Obviously, however, the theory breaks down when $\epsilon$ is too small. Fig \ref{fig:alpha} shows $\alpha_{xy}$ at 0.5 T in the 4 nm sample in a semi-log plot, across the critical temperature. One can see that It varies smoothly by 3 orders of magnitude, from a few A/K.m in the vortex liquid phase to 0.001 A/K.m. Above the critical temperature, when $\epsilon > 0.08$, the temperature dependence is $\varepsilon^{-1}$, as discussed above. Interestingly, the data does not show any discontinuity at the critical temperature. The smooth transition indicates an intimate link between the two regimes separated by the critical temperature. This is remarkable since  while there is a fairly quantitative understanding of the Nernst signal in the normal state (when $\epsilon$ is finite, the amplitude of the much larger Nernst response below the critical temperature (and associated with the vortex liquid) is yet to be quantitatively understood \cite{Sergeev_2010,Rischau2021,Sergeev2021,Behnia_2023,guo2024amplitudevortexentropysemiclassical}. 

We note that the amplitude of $\alpha_{xy}$ at low temperature approaches $2ek_B/h$, the ratio of the Boltzmann to the magnetic flux of a vortex $h/2e$, divided by the lattice constant (0.44 nm), or by $\xi_0^{Hc2}$, as indicated by the dashed horizontal lines. Taken at its face value, this observation indicates a vortex transport entropy per sheet of the order of the Boltzmann constant. Available data in cuprate superconductors \cite{HUEBENER-2021}, in $\kappa$-(BEDT-TTF)$_2$X organic superconductors \cite{Logvenov1997} and in Nb-doped strontium titanate \cite{Rischau2021} indicate a similar conclusion. Quantitative understanding of the amplitude of the vortex Nernst signal is a subject of ongoing research.   

In summary, we measured the resistivity and the Nernst coefficient of NbN thin films with a thickness ranging from 4 to 30 nm. We detected a Nernst response both below and above the superconducting critical temperature. We found that in the normal state, the amplitude of the Nernst conductivity depends on thickness as expected in a 3D picture, and displays a linear variation with reduced temperature as expected in a 2D picture. We argued that a 2+1D picture can reconcile the two features. The Nernst conductivity is continuous across the critical temperature, indicating an intimate link between the Nernst response by the short-lived Cooper pairs and the one by the superconducting vortices.

\section{Acknowledgments}
We thank Mikhail Feigel'man  and Christoph Strunk for discussions We acknowledge support from the \^Ile de France regional council and by the Agence Nationale de la Recherche (ANR-19-CE30-0014-04). B.F was supported by  Jeunes Equipes de l$'$Institut de Physique du Coll\`ege de France. 

\section{Data Availability}
The supporting data for this article are openly available from the PSL research data repository \cite{dataarticle}.

\bibliography{biblio}
\end{document}